\documentclass[aps,prd,superscriptaddress,showpacs,preprintnumbers,10pt]{revtex4}
\usepackage{graphicx,color,here}
\newcommand{\be}{\begin{equation}}
\newcommand{\ee}{\end{equation}}
\newcommand{\bea}{\begin{eqnarray}}
\newcommand{\eea}{\end{eqnarray}}

\def\lsim{\mathrel{\rlap{\lower4pt\hbox{\hskip1pt$\sim$}}\raise1pt\hbox{$<$}}}
\def\gsim{\mathrel{\rlap{\lower4pt\hbox{\hskip1pt$\sim$}}\raise1pt\hbox{$>$}}}
\def\nostrocostruttino#1\over#2{\mathrel{\mathop{\kern 0pt \rlap
{\hbox{$#1$}}} \hbox{\kern-.135em $#2$}}}

\def\Vec#1{{\bf #1}}

\def\D{{\mathrm d}}
\def\E{{\mathrm e}}
\begin{document}
%%%%%%%%%%%%%%%%%%%%%%%%%%%%%%%%%%%%%%%%%%%%%%%%%%%%%%%%%%%%%%%%%%%%%%%%%%%%%%
%\preprint{}
\title{Azimuthal asymmetries
in unpolarized Drell-Yan processes\\ and the Boer-Mulders 
distributions of antiquarks}

\author{Vincenzo Barone}
\affiliation{Di.S.T.A., Universit\`a del Piemonte Orientale
``A. Avogadro'', 15121 Alessandria, Italy \\ INFN, Gruppo Collegato di Alessandria, 15121
Alessandria, Italy}
\author{Stefano Melis}
\affiliation{Di.S.T.A., Universit\`a del Piemonte Orientale
``A. Avogadro'', 15121 Alessandria, Italy \\ INFN, Gruppo Collegato di Alessandria, 15121
Alessandria, Italy}
\author{Alexei Prokudin}
\affiliation{Jefferson Laboratory, 12000 Jefferson Avenue, 
Newport News, VA 23606}

%\date{}

\begin{abstract}

Using a previous extraction of the quark Boer-Mulders distributions 
from semiinclusive deep inelastic scattering data, we fit the 
unpolarized Drell-Yan data on the $\cos 2 \phi$ asymmetry, 
 determining the antiquark Boer-Mulders distributions. 
A good agreement with the 
data is found in the region of low $q_T$, where the 
transverse-momentum factorization approach applies.

\end{abstract}


\maketitle

\vspace{0.5cm}

\section{Introduction} 

One of the most relevant results of high-energy spin 
physics in the last decade has been the discovery of 
many interesting correlations between the transverse momentum 
and the transverse spin of quarks 
(for reviews, see Ref.~\cite{Barone:2001sp,Barone:2010}). A surprising consequence 
of these correlations is that there may exist non trivial 
spin effects in {\em unpolarized} hard processes,    
generated by a leading-twist, chiral-odd, transverse-momentum dependent
distribution function, the so called Boer--Mulders function 
$h_1^{\perp}(x, \Vec k_T^2)$ \cite{Boer:1997nt}, 
which represents a transverse--polarization asymmetry of quarks 
inside an unpolarized hadron. 
The origin of $h_1^{\perp}$ was clarified in 
Refs.~\cite{Collins:2002kn,Brodsky:2002cx,Brodsky:2002rv,Belitsky:2002sm}
and the first
calculation  in a realistic 
quark--diquark model was reported by Goldstein and Gamberg
\cite{Goldstein:2002vv}.
In 1999 Boer \cite{Boer:1999mm} suggested that 
$h_1^{\perp}$ could explain the large
$\cos 2\phi$ asymmetries
observed in unpolarized $\pi N$ 
Drell-Yan production \cite{Falciano:1986wk,Guanziroli:1987rp,Conway:1989fs}, 
which were not understood in terms 
of purely perturbative QCD effects 
\cite{Chiappetta:1986yg,Brandenburg:1993cj,Falciano:1986wk}. 
This finding was confirmed by more refined 
model calculations \cite{Lu:2004hu}. 
A comparable or even larger asymmetry 
is predicted for $p \bar p$ Drell-Yan 
production 
\cite{Boer:2002ju,Bianconi:2004wu,Sissakian:2005yp,Sissakian:2008th,Gamberg:2005ip,Barone:2006ws}, 
a process to be studied 
in the next years at the GSI High-Energy Storage 
Ring \cite{Kotulla:2004ii,Barone:2005pu}. 
While $\pi N$ and $p \bar p$ probe valence distributions, 
 $pp$ and $pD$ 
Drell-Yan reactions are sensitive to the sea distributions and 
therefore the corresponding $\cos 2 \phi$ asymmetries are 
expected to be smaller. This is indeed what the 
E866/NuSea experiment found: the $\cos 2 \phi$ 
dependence observed in $p D$ dimuon production is  
of the order of few percent \cite{Zhu:2006gx,Zhu:2008sj}.

A $\cos 2 \phi$ asymmetry also occurs 
in unpolarized semiinclusive deep inelastic scattering (SIDIS), 
where it has been measured in the low transverse-momentum 
region 
by HERMES \cite{Giordano:2009hi}, 
COMPASS \cite{Bressan:2009eu} and CLAS
\cite{Osipenko:2008rv}. 
In SIDIS the Boer--Mulders distribution 
couples to a chiral-odd fragmentation function, 
the Collins function $H_1^{\perp}$ \cite{Collins:1992kk}, 
which describes the fragmentation 
of transversely polarized quarks into polarized
hadrons. Recently, we presented 
a systematic phenomenological analysis of the various contributions to  
the $\cos 2 \phi$ asymmetries in unpolarized SIDIS \cite{Barone:2008tn}, 
and of the preliminary HERMES and COMPASS results
\cite{Barone:2009hw}. In the 
kinematics of these experiments  
the perturbative term is found to be negligible, whereas 
the order-$k_T^2/Q^2$ contribution from non-collinear 
kinematics
(the so-called Cahn effect \cite{Cahn:1978se,Cahn:1989yf}) 
is quite large. The 
Boer--Mulders effect is also sizable and generates
a negative (positive) asymmetry for $\pi^+$ ($\pi^-$), due to the expected 
negative sign of the $u$ distribution. 
Combining the Cahn contribution 
(which is positive and roughly the same for $\pi^+$ and $\pi^-$) 
with the Boer-Mulders contribution, we predicted a $\pi^-$ 
asymmetry larger than the $\pi^+$ asymmetry \cite{Barone:2008tn}.   
Our results turned out to be in fair agreement with the first SIDIS 
measurements of 
the $\cos 2 \phi$ asymmetry, 
as shown in our most recent study \cite{Barone:2009hw}, 
where we attempted an extraction 
of the Boer-Mulders distributions from the HERMES and COMPASS 
preliminary data. Since the present statistics 
is not sufficient to allow a complete determination 
of $h_1^{\perp}$, the strategy of Ref.~\cite{Barone:2009hw} 
was to use the same functional form as the Sivers function 
obtained from SIDIS data in Ref.~\cite{Anselmino:2008sga}
and parametrize the normalization coefficient. Results compatible 
with impact-parameter expectations \cite{Burkardt:2005hp}
combined with lattice findings \cite{Gockeler:2006zu}, 
and with model calculations \cite{Pasquini:2010af} were found. 
However, the  SIDIS data leave the Boer-Mulders sea 
totally unconstrained. Thus, in order to get the 
antiquark distributions $\bar h_1^{\perp}$ one needs 
an extra source of information.

The purpose of the present paper is indeed 
to extract some information on the antiquark Boer-Mulders 
distributions from the Drell-Yan $pp$ and $p D$ data 
on the $\cos 2 \phi$ 
asymmetry \cite{Zhu:2006gx,Zhu:2008sj}. We perform a fit to these data
in the region of low $q_T$, where the transverse-momentum  
factorization is expected to be valid \cite{Ji:2004xq} and the perturbative 
effects are small. Since in Drell-Yan processes the Cahn contribution 
to the 
$\cos 2 \phi$ asymmetry is known to be negligible \cite{Schweitzer:2010tt}, 
an analysis in terms of the Boer-Mulders effect alone is possible.

%%%%%%%%%%%%%%%%%%%%%%%%%%%%%%%%%%%%%%%%%%%%%%%%%%%%%%%%%%%%%%%%%

\section{The $\cos 2 \phi$ asymmetry in Drell-Yan processes}

The angular differential cross section for the unpolarized
Drell-Yan process is usually parametrized as
\be
 \frac{1}{\sigma^{\rm DY}}\frac{d \sigma^{\rm DY}}{d \Omega}=
\frac{3}{4\pi}\frac{1}{\lambda+3} \left (1+\lambda\cos^2\theta
 +\mu
\sin2\theta\cos\phi+\frac{\nu}{2}\sin^2\theta\cos2\phi\right )\,
.\label{angular}
\ee
where $\theta$ and $\phi$ are, respectively, the polar angle and
the azimuthal angle of dileptons in a dilepton 
center of mass frame. In particular, we adopt 
the Collins--Soper frame \cite{Collins:1977iv}, 
where $\theta$ is the angle between 
the dilepton axis and the bisector 
of $\Vec P_1$ and $- \Vec P_2$ (the momenta  
of the colliding hadrons), and $\phi$ 
is the angle between the lepton and hadron planes.
We denote by $q_T \equiv \vert \Vec q_T \vert$ the 
transverse momentum of the lepton pair (or, equivalently, 
of the virtual photon).  
 The $\nu$ parameter in (\ref{angular}) is the $\cos 2 \phi$ 
asymmetry we are interested in.

A non-collinear factorization
theorem for Drell-Yan process has been proven by Ji, Ma
and Yuan \cite{Ji:2004xq} for $q_{T} \ll Q$ (where $Q$ is 
the invariant mass of the lepton pair).  
At order $\alpha_s^0$, 
the $\phi$-independent term of the
unpolarized Drell-Yan cross section is
\be
 \frac{d\sigma^{\rm DY}}{d\Omega \, dx_1 \, d x_2 \,
d^2\Vec q_T} = \frac{\alpha^2_{\rm em}}{12 \, Q^2} (1 + \cos^2
\theta) \,  \sum_{a}e_a^2
\int d^2\Vec k_{1T} \, d^2\Vec k_{2T} \, 
\delta^2(\Vec k_{1T} +\Vec k_{2T}-\Vec q_T) \,  [f_1^a(x_1, k_{1T}^2)
\bar f_1^{a}(x_2, k_{2T}^2) + ( 1 \leftrightarrow 2)],\label{cs}.
\ee
Here  $f_1 (x, \Vec k_{1T}^2)$ is the
unintegrated quark number density. 

%We will consider the $\nu$ term in (\ref{angular}). It is known 
%that gluon radiation processes give rise to a non-zero 
%$\cos 2 \phi$ asymmetry, which
%in case of $q \bar q$ annihilation dominance is given by 
%$\nu = Q_T^2/(M^2 + 3 Q_T^2/2)$ \cite{Collins:1978yt}. 
%Various quantitative 
%analyses \cite{Chiappetta:1986yg,Brandenburg:1993cj,Falciano:1986wk} 
%show that perturbative corrections are unable to reproduce both the 
%magnitude and the $Q_T$-dependence of $\nu$ 
%as observed by NA10 in the region $M \sim 4-8$ GeV,  but 
%for lower $M$ values the perturbative asymmetry might be 
%relevant. 
%In the present paper we 
%will focus on the 
%Boer--Mulders contribution to the $\cos 2 \phi$ 
%asymmetry.

The Boer-Mulders
contribution to the unpolarized cross-section reads~\cite{Boer:1999mm}
\begin{eqnarray}
 \left. \frac{d\sigma^{\rm DY}}{d\Omega \, dx_1 \, d x_2
\, d^2\Vec q_T} \right \vert_{\cos 2 \phi} &=& \frac{\alpha^2_{\rm
em}}{12 \, Q^2} \sin^2 \theta 
\, \sum_a e_a^2\int d^2\Vec k_{1T}
d^2\Vec k_{2T} \, 
 \delta^2(\Vec k_{1T}+\Vec k_{2T}-\Vec q_T)
\nonumber \\
& &  \times 
\frac{(2\, \hat{\Vec
h}\cdot \Vec k_{1T} \hat{\Vec h}\cdot \Vec k_{2T}
-\Vec k_{1T}\cdot \Vec k_{2T})}{m_N^2} 
 \,[ h_1^{\perp a}(x_1, k_{1T}^2)\bar  h_1^{\perp a}(x_2,
k_{2T}^2)\cos2\phi  \; + \; ( 1 \leftrightarrow 2)] ,\label{bm}
\end{eqnarray}
with $\hat{\Vec h} \equiv \Vec q_T/q_T$. 

{From} Eqs.~(\ref{cs}) and (\ref{bm}) we get the following
expression for the coefficient $\nu$ (setting
$\lambda=1$, $\mu=0$) :
\begin{equation}
\nu=\frac{2\sum_{a}e_a^2\, \mathcal{H}[h_1^{\perp a}, \bar h_1^{\perp
a}]}{ \sum_{a}e_a^2\, \mathcal{F}[f_1^a, \bar f_1^{a}]}, \label{nu}
\end{equation}
with the following notations:
\begin{eqnarray}
 \mathcal{F}[f_1^a, \bar f_1^{a}] &=&  \int d^2\Vec k_{1T} 
\, d^2\Vec
k_{2T}\delta^2(\Vec k_{1T}+\Vec k_{2T}-\Vec q_T)  
\times f_1^a(x_1, k_{1T}^2) \bar f_1^{a}(x_2, k_{2T}^2)\nonumber\\
& =& 
\int \D k_{1T} \, k_{1T} \, \int_0^{2 \pi} \D \chi \, 
 f_1^{a}(x_1, k_{1T}^2) \, \bar f_1^{a}(x_2, \vert \Vec q_T -
\Vec k_{1T} \vert^2)\,,\label{fconv}
\end{eqnarray}
\begin{eqnarray}
 \mathcal{H}[h_1^{\perp a},\bar h_1^{\perp a}] &=& 
\, \int d^2\Vec k_{1T}
\, d^2\Vec k_{2T}\delta^2(\Vec k_{1T}+\Vec k_{2T}-\Vec q_T)
\, \frac{(2\hat{\Vec h}\cdot \Vec k_{1T} \hat{\Vec h}\cdot \Vec
k_{2T} -\Vec k_{1T}\cdot \Vec k_{2T})}{m_N^2} \, 
 h_1^{\perp a}(x_1, k_{1T}^2)
\bar h_1^{\perp a}(x_2,k_{2T}^2)
\nonumber \\
& =&  
\int \D k_{1T} \, k_{1T} \, \int_0^{2 \pi} \D \chi 
\,  \frac{
k_{1T}^2+ q_T\, k_{1T} \, \cos \chi - 2 \,  k_{1T}^2 \, \cos^2 \chi
}{m_N^2}  
 \, h_1^{\perp a}(x_1, k_{1T}^2) \, \bar h_1^{\perp a}(x_2,
\vert \Vec q_T - \Vec k_{1T} \vert^2)\, ,\label{hconv}
\end{eqnarray}
where $\chi$ is the angle between $\Vec q_T$ and $\Vec k_{1T}$ 
and we omitted for simplicity the $(1 \leftrightarrow 2)$ terms. 
 The
asymmetry $\nu$ in Eq.~(\ref{nu}) depends on the
kinematic variables $x_1$, $x_2$, $Q$ and $q_T$. 

In general, there is another contribution to $\nu$ 
arising from the Cahn effect, that is from purely kinematic 
transverse-momentum corrections to the ordinary 
parton model formulas. However, as shown in Ref.~\cite{Schweitzer:2010tt}, 
the Cahn contribution to $\nu$ is proportional to 
$(\langle k_{1T}^2 \rangle - \langle k_{2T}^2 \rangle )^2$ 
and hence negligible, or even strictly vanishing when 
the average transverse momenta of quarks (or antiquarks) 
in the two colliding hadrons are equal (which is what we assume 
here). Thus, the Boer-Mulders 
effect is the only non-perturbative source of a 
$\cos 2 \phi$ asymmetry up to order $q_T^2/Q^2$.

\section{Parametrizations of distribution and fragmentation functions}
Let us consider first of all 
the $k_T$-dependent unpolarized distribution functions (where 
$k_T$ stands either for $k_{1T}$ or 
for $k_{2T}$). We assume that  
these functions have a Gaussian behavior in $k_T$,  
\be
f_1^a (x, k_T^2) = f_1^a (x) \frac{\E^{-{ k_T^2}/
{\langle  k_T^2 \rangle}}}{\pi \langle  k_T^2 \rangle }\,
\ee
which is  
 supported by lattice studies \cite{Musch:2007ya} 
and by a recent phenomenological study of SIDIS and 
DY cross sections \cite{Schweitzer:2010tt}.   
The integrated unpolarized distribution 
functions $f_1^q$ are taken
from the GRV98 fit \cite{Gluck:1998xa}.

Since the  available SIDIS data
do not allow  a full extraction of the Boer-Mulders function, 
in Ref.~\cite{Barone:2009hw} we assumed
$h_1^{\perp}$ to be simply proportional 
to the Sivers function $f_{1T}^{\perp}$, 
 \be
h_1^{\perp a}(x, k_T^2) = \lambda_a \,  
f_{1T}^{\perp a}(x, k_T^2)\,, 
\label{lambda}
\ee 
with $f_{1T}^{\perp a}$ taken from a
phenomenological analysis of the Sivers asymmetry \cite{Anselmino:2008sga}
and the coefficient $\lambda_a$ fitted to the SIDIS $\cos 2 \phi$ data.
Various theoretical arguments (based 
on the impact-parameter picture \cite{Burkardt:2005hp}, on 
large-$N_c$ arguments \cite{Pobylitsa:2003ty}, and on model calculations 
\cite{Pasquini:2008ax,Courtoy:2009pc}) 
suggest that the $u$ and $d$ components 
of $h_1^{\perp}$, at variance with $f_{1T}^{\perp}$, 
 should have the same sign and in particular 
be both negative (which means that $\lambda_d$ should be 
negative). This is indeed what we found in Ref.~\cite{Barone:2009hw}.  
Moreover, the impact-parameter approach \cite{Burkardt:2005hp} combined 
with lattice results \cite{Gockeler:2006zu} predicts 
a $u$ component of $h_1^{\perp}$ larger in magnitude than 
the corresponding component of $f_{1T}^{\perp}$, and the  
$d$ components of $h_1^{\perp}$ and $f_{1T}^{\perp}$ with approximately 
the same magnitude (and opposite sign). 

In our SIDIS analysis the quark Boer-Mulders distributions
were parametrized as 
\be
h_1^{\perp a}(x, k_T^2) = N_a \, x^{\alpha_a} (1 - x)^{\beta_a}
\, \E^{- k_T^2/\mu^2} \, f_1^a (x, k_T^2)
\ee
with the parameters $\alpha_a, \beta_a, \mu$ borrowed from  
Ref.~\cite{Anselmino:2008sga} and $N_a$ fitted to the data. 
The values of these parameters are 
collected in  Table~\ref{fitpar}. Notice that, when used to 
calculate DY asymmetries, the Boer-Mulders distributions 
determined in SIDIS must be sign-reversed \cite{Collins:2002kn,Brodsky:2002rv}.
\begin{table}[t]
\begin{center}
\begin{tabular}{|l l|}
\hline
~~~$N_{u} =- 18$ \hspace*{1cm} &
~~~$N_{d} = - 45$ \\
~~~$\alpha _u = 0.73$ & 
~~~$\alpha_d = 1.08$  \\
~~~$\beta_u= \beta_d \;\;= 3.46$ &
~~~$\mu^2 = 0.34$ GeV$^2$~   \\
\hline
\end{tabular}
\end{center}
\caption{Parameters of the Boer-Mulders quark distributions
\cite{Barone:2009hw}.}
\label{fitpar}
\end{table}

Concerning the Boer-Mulders antiquark distributions,  
the SIDIS data are quite insensitive to them. Therefore, 
 in Ref.~\cite{Barone:2009hw}
these distributions were not fitted to the data, but just taken to be 
equal in magnitude to the corresponding Sivers distributions. 
 
The Drell-Yan process, on the contrary, probes 
the products of quark and antiquark distributions.  
Thus the Drell-Yan measurements 
of the $\cos 2 \phi$ asymmetry \cite{Zhu:2006gx,Zhu:2008sj} 
can at least in principle  
give information about the antiquark sector 
of the Boer-Mulders function. An analysis of 
the DY $\cos 2 \phi$ asymmetry has been already 
performed by other authors \cite{Zhang:2008nu,Lu:2009ip}. However, when 
extracting the Boer-Mulders distributions from the present 
DY data, one should keep in mind that  
the $k_T$-factorization approach applies to the 
low-$q_T$ region only, whereas at large $q_T$ 
the observed $\nu$ values  are likely to be explainable in terms 
of perturbative QCD \cite{Boer:2006eq,Berger:2007jw}.  
Therefore, in our analysis we will only consider 
the low-$q_T$ DY data.

The antiquark Boer-Mulders distributions are parametrized as 
\be
\bar{h}_1^{\perp a}(x, k_T^2) = N_{\bar a} \, x^{\alpha_{\bar a}} 
(1 - x)^{\beta_{\bar a}}
\, \E^{- k_T^2/\mu^2} \, \bar{f}_1^a (x, k_T^2)\,, 
\ee
with the same $\alpha, \beta$ and $\mu$ parameters as for 
the Sivers antiquark distributions, 
and the normalization coefficients $N_{\bar u}$ and $N_{\bar d}$ 
fitted to the data.

The final parameters to be considered are the   
average values of the quark and antiquark transverse 
momenta in the two hadrons,  $\langle k_{1T}^2 \rangle$ and $\langle 
k_{2T}^2 \rangle$. We take these parameters to be equal to each 
other, $\langle k_T^2 \rangle 
\equiv \langle k_{1T}^2 \rangle = \langle 
k_{2T}^2 \rangle$, and we choose for them two different values: 
\be
{\rm Fit} \; 1: \; 
\langle k_{T}^2 \rangle
= 0.25 \; {\rm GeV}^2\,; 
\;\;\;
{\rm Fit} \; 2: \; 
\langle k_{T}^2 \rangle  
= 0.64 \; {\rm GeV}^2\,. 
\label{kTvalues}
\ee
The smaller value, 0.25 GeV$^2$, is the one we used in our analysis 
of the Boer-Mulders effect in SIDIS
\cite{Barone:2009hw}, and is taken from the  
study of the Cahn effect 
of Ref.~\cite{Anselmino:2005nn}. 

The larger value, 0.64 GeV$^2$, is the one obtained 
by D'Alesio and Murgia
\cite{D'Alesio:2004up} in their analysis 
of $pp$ scattering data at $\sqrt{s} \simeq 20$ GeV. 
A recent phenomenological study of the transverse-momentum 
dependence of DY cross sections has obtained 
a very similar value \cite{Schweitzer:2010tt}.

\section{Results}

The E866/NuSea Collaboration presented data on the angular 
distributions of DY dimuons in $pD$ 
\cite{Zhu:2006gx} and $pp$ interactions \cite{Zhu:2008sj} 
over the kinematic range: 
\[
4.5 < Q < 15 \; {\rm GeV}\,, \;\;\;
0< q_T < 4 \; {\rm GeV}\,, \;\;\;
0<x_F <0.8\,. 
\]
Concerning $q_T$, as already mentioned, we will only consider 
the data in the $q_T < 1.5 $ GeV region, where the $k_T$ 
factorization applies and the perturbative contribution, 
roughly proportional to $q_T^2/Q^2$, is small.

%%%%%%%%%%%%%%%%%%%%%%%%%%%%%%%%%%%%%%%%%%%%%%%%%%%%%%%%%%%%%%%%%%%%%%%%%%%%%%%
\begin{figure}[t]
\hspace{-1cm}
\includegraphics[width=0.35\textwidth, angle=-90]
{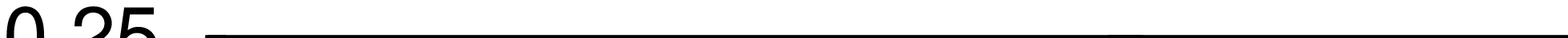}

\caption{\label{fig:nupD}
 The curves represent the 
results of our fit 1 to the $\nu$ asymmetry of $pD$ DY production. Data 
are from Ref.~\cite{Zhu:2006gx}.}
\end{figure}
%%%%%%%%%%%%%%%%%%%%%%%%%%%%%%%%%%%%%%%%%%%%%%%%%%%%%%%%%%%%%%%%%%%%%%%%%%%%%%%

\begin{figure}[thb]
\hspace{-1cm}
\includegraphics[width=0.35\textwidth, angle=-90]
{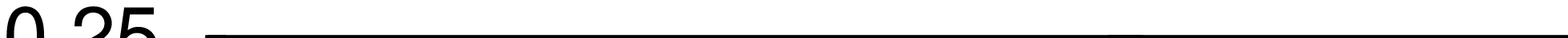}

\caption{\label{fig:nupp}
Our fits to the $\nu$ asymmetry of $pp$ DY production. Data 
are from Ref.~\cite{Zhu:2008sj}.}
\end{figure}

Our fit 1 to the E866/NuSea data on $\nu$ is shown 
in Fig.~\ref{fig:nupD} for $pD$ and in Fig.~\ref{fig:nupp} 
for $pp$. 
The 
$\chi^2$ per degree of freedom is 1.24.
The vertical error bars represent the statistical uncertainties. 
Notice that the experimental bins in all variables ($x_1$, $x_2$ and 
$q_T$) are quite large: for instance, the three points 
in the $q_T$ plot correspond (in GeV) to (0,0.5), (0.5,1.0), (1.0,1.5). 
In all figures we only showed the central values of the bins.  

Concerning fit 2,  
the corresponding curves overlap those of fit 1, with the same 
value of $\chi^2$/d.o.f. The difference between the Gaussian widths 
of the distribution is in fact compensated by the different normalizations 
of the antiquark distributions that we obtain from the two fits.

%%%%%%%%%%%%%%%%%%%%%%%%%%%%%%%%%%%%%%%%%%%%%%%%%%%%%%%%%%%%%%%%%%%%%%%%%%%%%%%
%%%%%%%%%%%%%%%%%%%%%%%%%%%%%%%%%%%%%%%%%%%%%%%%%%%%%%%%%%%%%%%%%%%%%%%%%%%%%%%

\begin{figure}[thb]
\hspace{-1cm}
\includegraphics[width=0.35\textwidth, angle=-90]
{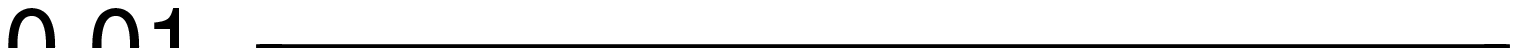}

\caption{\label{fig:bm}
The first $k_T^2$ moments of the antiquark distributions 
from Fit 1 (solid curves) and Fit 2 (dashed curves).}
\end{figure}

The values of the parameters of the antiquark distributions are 
listed in Table~\ref{fitpar2}.  The first $k_T^2$ moments 
of the antiquark distributions, i.e. 
\[
\bar{h}_{1}^{\perp (1)}(x) = \int d^2 \Vec k_T \, 
\frac{k_T^2}{2 m_N^2} \, \bar{h}_1^{\perp} (x, k_T^2)
\] 
 are plotted in Fig.~\ref{fig:bm}. One may notice the   
sensible difference between the distributions extracted 
from the two fits. The width of the Gaussian distribution 
is clearly a crucial parameter, but the scarcity of present 
data does not allow extracting it from the fit. 
A combined analysis of SIDIS and DY data, and possibly 
of various azimuthal asymmetries, might help to improve 
the situation.

\begin{table}[t]
\begin{center}
\begin{tabular}{|c|c|c|c|}
\hline
\multicolumn{4}{|r|}{$\alpha _{\bar u} = \alpha_{\bar d} = 0.79$,  
$\beta_{\bar u} = \beta_{\bar d}= 3.46 $,  
$\mu^2 = 0.34$ GeV$^2$} \\
\hline
Fit 1 &
$N_{\bar u} = 3.6 \pm 1.0$  &
$N_{\bar d} = 1.7 \pm 1.4$ & 
$ \langle k_T^2 \rangle = 0.25$ GeV$^2$ \\
\hline
Fit 2 &
$N_{\bar u} = 6.4 \pm 1.7$  &
$N_{\bar d} = 3.0 \pm 2.4 $ &
$ \langle k_T^2 \rangle = 0.64$ GeV$^2$ \\
\hline
\end{tabular}
\end{center}
\caption{Parameters of the Boer-Mulders antiquark distributions.}
\label{fitpar2}
\end{table}
%

%%%%%%%%%%%%%%%%%%%%%%%%%%%%%%%%%%%%%%%%%%%%%%%%%%%%%%%%%%%%%%%%%%%%%%%%%%%%%%%
%%%%%%%%%%%%%%%%%%%%%%%%%%%%%%%%%%%%%%%%%%%%%%%%%%%%%%%%%%%%%%%%%%%%%%%%%%%%%%%

Finally, let us compare the present work with with the results of 
a recent analysis \cite{Zhu:2008sj} of the E866/NuSea measurements.  
In Ref.~\cite{Zhu:2008sj} both the quark and the antiquark distributions 
have been extracted from $pp$ and $pD$ DY data. Since in the 
DY cross section $h_1^{\perp a}$ always couples to 
$\bar{h}_1^{\perp a}$, only the magnitude of the products 
$h_1^{\perp a} \bar{h}_1^{\perp a}$ is actually determined, 
and the signs of the distributions are undefined. In our 
analysis the use of both the DY and the SIDIS data allows 
constraining separately the magnitudes and the signs 
of the Boer-Mulders quark and antiquark distributions.

\section{Conclusions}

We performed a fit to the DY data on the $\cos 2\phi$ 
asymmetry, showing that using the Boer-Mulders 
quark distributions previously extracted from 
SIDIS and a new set of antiquark distributions
one can obtain a reasonably good description 
of $\nu$. However, the resulting Boer-Mulders 
functions depend rather strongly on the width of the 
Gaussian distributions, that is on the 
average $k_T^2$. In order to achieve a 
more precise determination of $h_1^{\perp}$ 
for quarks and antiquarks a combined fit of SIDIS 
and DY data, and of $\cos \phi$ and $\cos 2 \phi$ 
asymmetries, must be performed. This work is now in 
progress.

\begin{acknowledgments}
We acknowledge support by the European Community - Research Infrastructure
Activity under the FP6 Program ``Structuring the European Research Area''
(HadronPhysics, contract number RII3-CT-2004-506078), by the Italian 
Ministry of Education, University and Research (PRIN 2008) and  
by the Helmholtz Association through
funds provided to the virtual institute ``Spin and Strong QCD''(VH-VI-231).
The work of one of us (S.M.) is also supported by 
Regione Piemonte. 
Authored by a Jefferson Science Associate, LLC under U.S. DOE Contract 
No. DE-AC05-06OR23177. The U.S. Government retains a non-exclusive, 
paid-up, irrevocable, 
world-wide license to publish
or reproduce this manuscript for U.S. Government purposes.
\end{acknowledgments}

%\bibliographystyle{h-physrev3.bst}
%\bibliography{biblio_costwophi}

\end{document}